\documentclass[%
 reprint,
 amsmath,amssymb,
 aps,
]{revtex4-2}

\usepackage{graphicx}
\usepackage{dcolumn}
\usepackage{bm}

\usepackage{graphicx}
\usepackage{amsmath}
\usepackage{epsfig}
\usepackage{amsmath,amsfonts,amssymb}
\usepackage[font=small,labelfont=bf, justification=justified,format=plain]{caption} 
\usepackage{times}
\usepackage{epsfig}
\usepackage{epstopdf}
\usepackage[colorlinks]{hyperref}

\usepackage{subcaption}
\usepackage[export]{adjustbox}

\usepackage{color}

\graphicspath{{Figs/}}

\begin{document}
\baselineskip=12pt
\def\black{\textcolor{black}}
\def\red{\textcolor{black}}
\def\blue{\textcolor{blue}}
\def\green{\textcolor{black}}
\def\be{\begin{equation}}
\def\ee{\end{equation}}
\def\bea{\begin{eqnarray}}
\def\eea{\end{eqnarray}}
\def\orc{\Omega_{r_c}}
\def\om{\Omega_{\text{m}}}
\def\E{{\rm e}}
\def\bearst{\begin{eqnarray*}}
\def\eearst{\end{eqnarray*}}
\def\peleven{\parbox{11cm}}
\def\peffec{\peight{\bearst\eearst}\hfill\peleven}
\def\pspace{\peight{\bearst\eearst}\hfill}
\def\ptwelve{\parbox{12cm}}
\def\peight{\parbox{8mm}}

\title{Cosmic velocity, density and halo mass function: Insights from deep learning}

\author{Saba \surname{Etezad-Razavi}$^{1}$}
\author{Erfan \surname{Abbasgholinejad}$^{1}$}
\author{Mohammad-Hadi \surname{Sotoudeh}$^{1,2,3}$}
\author{Farbod \surname{Hassani}$^{4}$}
\author{Sadegh \surname{Raeisi}$^{1}$}
\author{Shant \surname{Baghram}$^{1,5}$}
\email{baghram@sharif.edu}


%
%




\affiliation{
$^1$Department of Physics, Sharif University of Technology, Tehran 11155-9161, Iran\\
$^2$Department of Physics, Univesité de Montréal, Montréal, Canada\\
$^3$Mila - Quebec Artificial Intelligence Institute, Montréal, Canada\\
$^4$Institute of Theoretical Astrophysics, University of Oslo, 0315 Oslo, Norway \\
$^5$Research Center for High Energy Physics, Department of Physics, Sharif University of Technology, Tehran 11155-9161, Iran
}

\begin{abstract}
{{We discuss an implementation of a deep learning framework to gain insight into dark matter (DM) structure formation. We investigate the contribution of velocity and density field information to the construction of the halo mass function (HMF) in cosmological N-body simulations.
 We train a Convolutional Neural Network (CNN) on the initial snapshot of a DM-only simulation to predict the halo mass that individual particles fall into at $z = 0$, in the halo mass range of $10.5 < \log (M/M_{\odot}) < 14$.
We show that for the standard $\Lambda$CDM cosmology with amplitude of initial perturbations $A_s = 2 \times 10^{-9}$, the initial velocity and density fields have equivalent information, as expected in the linear regime, and manifest the power of our CNN to diagnose the redundant information.
To investigate the non-linear effects, we increase the initial power spectrum. In the linear regime, this is equivalent to decreasing the initial redshift.
 The CNN model trained on the simulation snapshots with large $A_s$ shows a considerable improvement in the HMF prediction when adding the velocity field information. Using our CNN map without further physical assumptions, we precisely evaluate when these non-linear effects become vital. Eventually, for the simulation with $A_s = 8 \times10^{-8}$, the model trained with only density information shows at least an $80\%$ increase in the mean squared error relative to the model with both velocity and density information. Our work shows the interpretability and ability of CNNs to read higher-order information from simple images, making them an excellent tool for cosmological studies.}}

\end{abstract}


\maketitle

\section{Introduction}\label{sec1}
The standard model of cosmology known as $\Lambda$CDM (cosmological constant plus cold dark matter) is established in the last two decades through observation of the cosmic microwave background radiation (CMB) \cite{Planck:2018vyg} and large-scale structure (LSS) surveys \cite{SDSS:2009ocz,BOSS:2016wmc,DES:2018csk}. Despite many successes of this model in describing the evolution of the Universe, and the statistical distribution of structures, the nature of dark energy, dark matter (DM) and the physics of the early Universe is still unknown. The future LSS surveys, for example, the Vera Rubin observatory, Euclid satellite, and  Nancy Groce Roman mission are planned to map the Universe in more detail over a wider range of redshifts \cite{Rhodes:2019lur}.
Accordingly, the study of LSS in the Universe employing high precision $N$-body simulations becomes crucial, especially to compare different models against precise observational data.
In the standard picture of structure formation, the galaxies, group of galaxies and galaxy clusters reside in DM halos \cite{White:1977jf}. These DM halos are the nodes of the cosmic web, which are structured by hierarchical formation. In this sense, the study of DM halo formation is essential for understanding galaxy formation and evolution and testing our cosmological models. The evolution of DM halos from an initial condition to the late time structures is a complicated non-linear process based on gravitational instability, mainly through mass accretion and mergers of DM structures \cite{Peebles1980}. Also, the knowledge of complicated baryon physics becomes essential to populate DM halos.
In this direction cosmological $N$-body simulations are used to model these non-linear processes \cite{Springel:2000yr,Springel:2005nw,Kuhlen:2012ft}.
On the other hand, the analytical models can give insights into the process of structure formation. One of the old but enlightened ideas of these models of structure formation is related to the nominal work of Press and Schechter (PS) \cite{Press:1973iz}. They propose that by looking at the statistics of the density field in higher redshifts, where the perturbations are almost Gaussian and linear, we can predict the resulting halo mass function by linear evolution of the density contrast field to the late time  (As a specific application of PS idea see\cite{Fard:2017oex}). This raises the question that what are the main features of the early Universe which are important to predict the DM halo mass function. The two ideas of peak theory \cite{Bardeen:1985tr} and excursion set theory (EST) \cite{Bond:1990iw} are theoretical developments to use the idea of using initial conditions to predict the late time distribution of matter. The two main features of these analytical models are the collapse model (spherical collapse, ellipsoidal collapse \cite{Sheth:1999su,Sheth:2001dp} and the smoothing functions of the initial density field \citep{Nikakhtar:2018qqg}.
Regardless of the massive success of analytical frameworks and simulation in shedding light on our understanding of structure formation, they still are not complete. Analytical models fail to accurately predict the formation of structures for the smallest halo ranges, and the simulations lack the physical insight that we could get from analytical models and leaves unanswered questions like, Is the initial information enough to predict halo mass function (HMF) accurately? What are the main features of the initial cosmic fields affecting the halo formation?

It is worth mentioning, that due to the enormous amount of data going to be delivered in near future from the cosmological surveys and simulations, we have already entered an era of big data in cosmology. Thus, the Machine Learning (ML) techniques become a complementary way to study the LSS of the Universe \citep{He:2018ggn}. ML can be used to have a more precise prediction for the LSS observables and to shed light on the physics and process of structure formation \citep{Lucie-Smith:2019hdl,Lucie-Smith:2020ris}. \\
In this work, we use ML methods to investigate the proposition of the analytical models of structure formation in predicting the late-time distribution of matter. Mainly, we aim to investigate the ideas of spherical collapse and ellipsoidal collapse.  The first one is related to the initial smoothed density and the second one incorporates the physics of velocity field and tidal shear of the collapse region. \\
{{
Our results, derived from a physical-model-independent procedure, confirm the well-established understanding that velocity information is crucial for accurately predicting the halo mass function (HMF) in the non-linear regime.
In the context of the PS, our goal is to show the possibility of choosing the initial simulation box at later times (lower redshifts) to predict the HMF. Nevertheless, to achieve this objective, it is important that we possess knowledge of the velocity field of the DM particles, as it significantly enhances the precision of our predictions. }}

We use Convolutional Neural Networks (CNN) to study the effect of density contrast and velocity field in LSS formation.
The organization of the paper is as follows: In Section\ref{sec2} we review the theoretical background of this work, and in Section \ref{sec3} we discuss the $N$-body simulation setups and the data being used in our work, namely we describe our simulations. Section\ref{sec4} is devoted to the technical description of our method and the deep learning framework. We present our results and interpretations in section \ref{sec5}, and finally, we have our conclusion and future remarks in section \ref{sec6}.
\section{Theoretical Background}
\label{sec2}
\begin{figure}
\includegraphics[width=\linewidth]{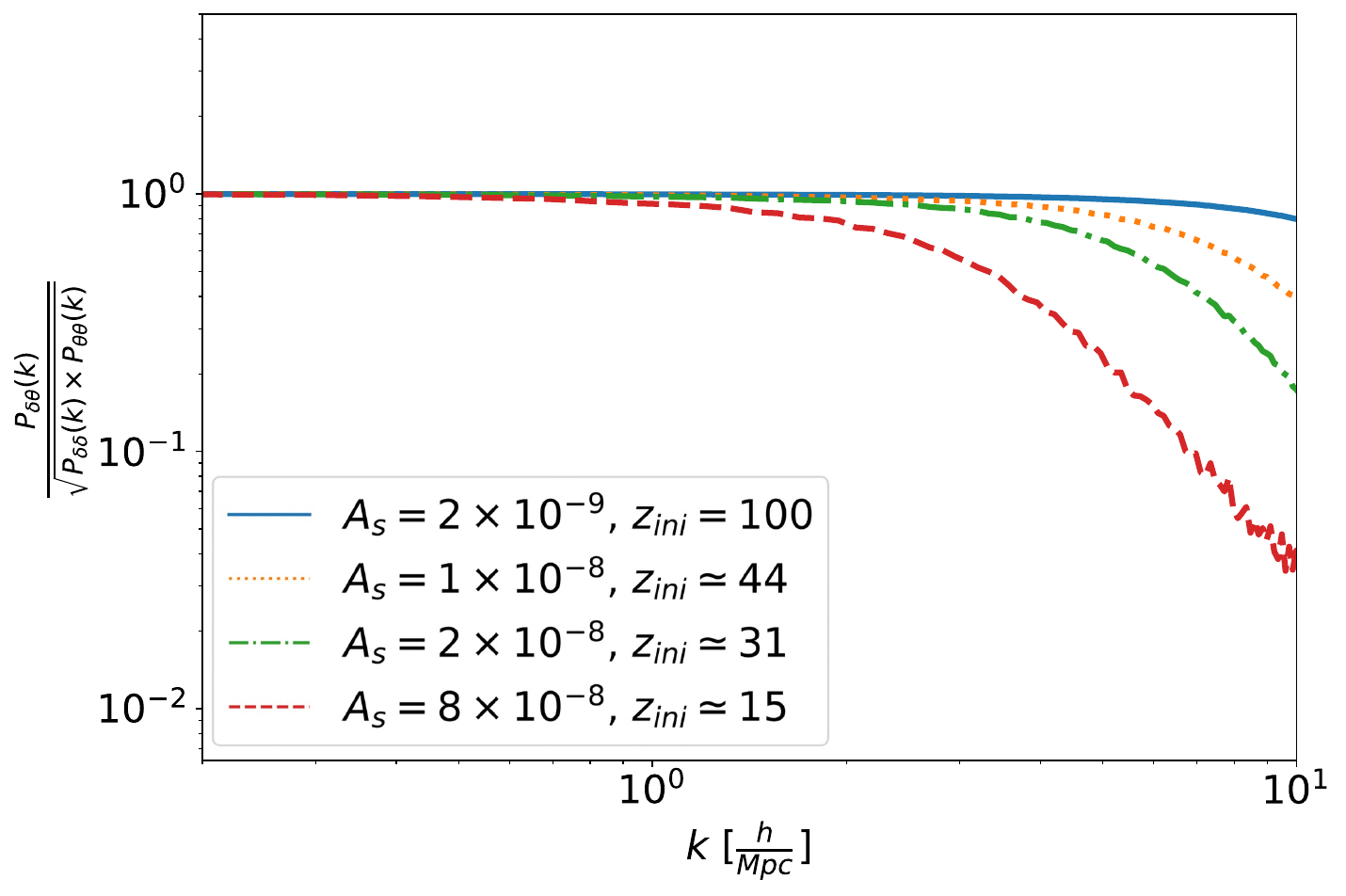}
\caption{The normalized cross power spectrum (NCP) is plotted versus wavenumber for different values of $A_s$. Each $A_s$ corresponds to a specific initial redshift. The unity shows that all information in the velocity field is present in the density field as well. The deviation from unity shows that the non-linear effect becomes important.}
\label{fig:normalized-cross}
\end{figure}

In the context of the standard model, DM halos are the host of the luminous matter \citep{Press:1973iz,Bond:1990iw}. Accordingly, the statistics and distribution of galaxies are promising tracers of DM distribution in the Universe \cite{Sheth:1999mn}.
There are two different approaches to studying the formation and evolution of DM halos, first through $N$-body simulations and second, utilizing theoretical ideas to study the mass profile of DM halos. 
The analytical models are mainly based on the idea of PS \citep{Press:1973iz}. In this framework, we study the Universe in high redshift where the perturbations are almost Gaussian and linear. Then, we predict the number density of DM halos by integrating the probability distribution (PDF) of density contrast from the critical density of collapse in spherical model $\delta_\text{sc} \simeq 1.69$ to infinity \citep{Press:1973iz}. The PDF is constructed using a specific smoothing scale. The mass enclosed in the smoothing region is equal to the mass of the DM halo formed later during the evolution of the overdense region.
Then the two ideas of peak theory \citep{Bardeen:1985tr} and excursion set theory \citep{Bond:1990iw} were developed accordingly. An important ingredient of these frameworks is the collapse model. The spherical collapse model \citep{Gunn:1972sv} only depends on the value of the over-density in the smoothed scale. However, a more realistic model of ellipsoidal collapse \citep{Sheth:2001dp} indicates the importance of a strong shearing field of velocity which makes the collapse of smaller mass halos more difficult than the larger halo mass. This effect manifests itself in a mass-dependent collapse barrier. To quantify the process we show how the non-linear distribution of DM structures is related to the initial condition.
The initial power spectrum of curvature perturbation is defined as
\begin{equation}
{\cal{P}}_{{\cal{R}}}=A_s (\frac{k}{k_p})^{n_s-1},
\end{equation}
where $A_s$ is the amplitude of the curvature perturbation, $n_s$ is the spectral index and $k_p$ is the pivot wavenumber. The amplitude and the spectral index are fixed by the CMB observations to the values $A_s\simeq 2\times 10^{-9}$ and $n_s \simeq 0.96$ \citep{Planck:2018vyg}.
We can find the matter power spectrum via the evolution of the scalar Bardeen potential through cosmic time for different modes as \citep{Dodelson:2003ft}
\begin{equation}
P_m(k,z) = A_{l} k^{n_s}T^2(k)D^2(z),
\end{equation}
where $A_{l}$ is the late time matter power spectrum amplitude, $T(k)$ and $D(z)$ are transfer function and growth function, respectively. The late time amplitude of matter power spectrum is related to the initial curvature amplitude as $A_l= A_s \times (8\pi^2k_p^{1-n_s})/({25\Omega_m^2H_0^4})$. \\
{{The growth function is the solution of the equation below, which results from the continuity, Euler and Poisson equations to find out the dynamics of dark matter density contrast defined $\delta_m = \rho_m/\bar{\rho}_m-1$, where $\bar{\rho}_m$ is the mean density of matter:
\be \label{eq:growth-diff}
\ddot{\delta}_m+2H\dot{\delta}_m - 4\pi G\rho_m \delta_m = 0,
\ee
where  $D(z) = \delta_m / \delta_m (a=1) $ and dot represent the derivative with respect to cosmic time. $\delta_m (a=1)$ is the density contrast at present time, where $a\equiv (1+z)^{-1}$ stands for scale factor normalized to unity at present time. The solution of the equation(\ref{eq:growth-diff})  for deep dark matter dominated era is $D(a) \propto a$. Accordingly the growth function is a monotonically increasing function with respect to scale factor. }}
We can obtain the scale of non-linearity which is a function of  redshift and mass  using the definition of variance $S$
\begin{equation} \label{eq:variance}
S(M,z) \equiv \sigma^2(M,z)=\int \frac{dk}{2\pi^2}k^2 P_m(k,z)\tilde{W}^2(kR),
\end{equation}
where $\tilde{W}(kR)$ is the Fourier transfer of the top-hat window function with smoothing scale $R$ related to the mass scale $M$.
For the standard cold dark matter model, the variance is a monotonic decreasing function of mass. Larger masses corresponds to lower variance scales. This means that by increasing the variance, we will probe the smaller mass scales. In this work, we increase the primordial curvature perturbation, which results in a higher variance mimicking the small mass range. \\
{{On the other hand, equation (\ref{eq:variance}), shows that the amplitude of the variance is also related to the growth function. It means that increasing the primordial power spectrum can be mimicked by studying the distribution of matter in later times where the growth function is larger. In cosmic history, the evolution of structures and the gravitational instability resulted in the emergence of non-linearity. In this case, the information on velocities becomes vital.}}
We will discuss this idea in the simulation section as well.
The number density of dark matter halos $n(M,z)$ can be expressed in terms of variance and the collapse threshold as \citep{Sheth:2001dp}
\begin{equation}
n(M,z) = \frac{\bar{\rho}_m}{M^2} f(\nu) \frac{d\ln \nu}{d\ln M},
\end{equation}
where $\bar{\rho}_m$ is the mean density of matter in the present time, $\nu = (\delta_\text{sc}/\sigma(M))^2$  is the height function and $f(\nu)$ is a universal function of $\nu$, related to the first up-crossing statistics in the EST \citep{Nikakhtar:2016bju}. For the ellipsoidal collapse, the universality function is given
\begin{equation} \label{eq:nuST}
\nu f(\nu) = \bar{A} [1+(\bar{a}\nu)^{-p}](\frac{{\bar a}\nu}{2})^{1/2}\frac{e^{-\bar{a}\nu/2}}{\sqrt{\pi}},
\end{equation}
where $p=0.3$, $\bar{a}=0.7$ and $\bar{A}$ is the normalization factor. The DM halo number density obtained from the universality function defined in equation (\ref{eq:nuST}) is known as the Sheth-Tormen (ST) number density. In the case of $\bar{a}=1$ and $p=0$, we will find the extended Press-Schechter mass function for spherical collapse \cite{Bond:1990iw}.
Note that the ratio of critical density in ellipsoidal collapse $\delta_\text{ell}$ to spherical collapse $\delta_\text{sc}$ model is defined in ST model as below \cite{Sheth:2001dp}:
\begin{equation} \label{eq:ellip-barrier}
\frac{\delta_\text{ell}}{\delta_\text{sc}} = \sqrt{\bar{a}} [1+\beta (\bar{a}\nu)^{-\alpha}],
\end{equation}
where $\bar{a}\approx 0.7$, $\alpha \approx 0.615$ and $\beta \approx 0.485$.
The halo mass function we used in this work for comparison with $N$-body simulation data is from \cite{Tinker:2008ff}, which calibrates the mass function at $z=0$ for virial masses in the range $10^{11} h^{-1} M_{\odot} \leq M \leq 10^{15} h^{-1} M_{\odot}$ to $5\%$:
\begin{equation}
\frac{dn}{dM}=f(\sigma)\frac{\bar{\rho}_m}{M}\frac{d\ln\sigma ^{-1}}{dM},
\end{equation}
where
\begin{equation}
f(\sigma)=A\left[(\frac{\sigma}{b})^{-a}+1\right]e^{-c/\sigma^2},
\end{equation}

with the fitting functions for constants 
\begin{eqnarray}
A&=& 
\begin{cases} 
0.1(\log \Delta) - 0.05     ~~~ &\Delta < 1600, \\ \nonumber
0.26                         ~~~ &\Delta \geq 1600,
\end{cases} \\
a&=&1.43+(\log\Delta - 2.3)^{1.5},\\  \nonumber
b&=&1.0+(\log\Delta - 1.6)^{-1.5},\\ \nonumber
c&=&1.2+(\log\Delta - 2.35)^{1.6},  \nonumber
\end{eqnarray}
where all logarithms are base 10 and $\Delta$ is the over-density, where in this work, we set it equal to $200$.
An important question to ask: what is the crucial information in the initial condition which helps to predict the distribution of dark matter? In linear perturbation theory through the continuity and Euler equations, there is a relation between the density contrast field and velocity. However, this assumption can be checked by studying the normalized cross-power spectrum of velocity and density contrast field.  
We define the normalized cross-power spectrum $\text{NCP}$ as
\begin{equation}
\rm{NCP}= \frac{P_{\delta\theta}(k)}{\sqrt{P_{\delta\delta}(k) P_{\theta\theta}(k)}},
\end{equation}
where $P_{\delta\delta}(k)$, $P_{\theta\theta}(k)$ are the density and momentum $( p = \rho V)$ divergence power-spectra and $\theta$ is defined as $\nabla . (1+\delta) V$. Using the momentum field instead of the velocity field is to avoid complexities rises in defining a velocity field in simulation voxels without any particle.  $P_{\delta\theta}(k)$ is the cross power spectrum of density and momentum divergence. In a linear regime by definition, we have $\text{NCP}=1$, which means that all the information is in the density-contrast field and the velocity field information is entirely constructible from the density field using the Poisson equation.
However, Fig.\ref{fig:normalized-cross}, shows a deviation from unity in small scales, as well as when we're increasing $A_s$ values. It means that in N-body simulations, because of non-linear effects, we have access to additional information in the velocity field. This is a point we use throughout this work to justify our results.
{{In Fig.\ref{fig:normalized-cross}, we plot the NCP for different values of initial conditions. It shows that by increasing the amplitude the scale of non-linearity goes to larger scales and we have more deviation from unity. Also, we have an equivalent interpretation, by increasing the amplitude of initial perturbation, the situation is similar to look at the perturbation in later times. In Fig.\ref{fig:normalized-cross}, we show that each amplitude corresponds to a specific initial redshift. From equation (\ref{eq:variance}), we have $D(z_2)/D(z_1)=\sqrt{A_2/A_1}$, where $z_1=100$ with $A_1=2\times 10^{-9}$.  }}\\
\begin{figure}
\includegraphics[width=\linewidth]{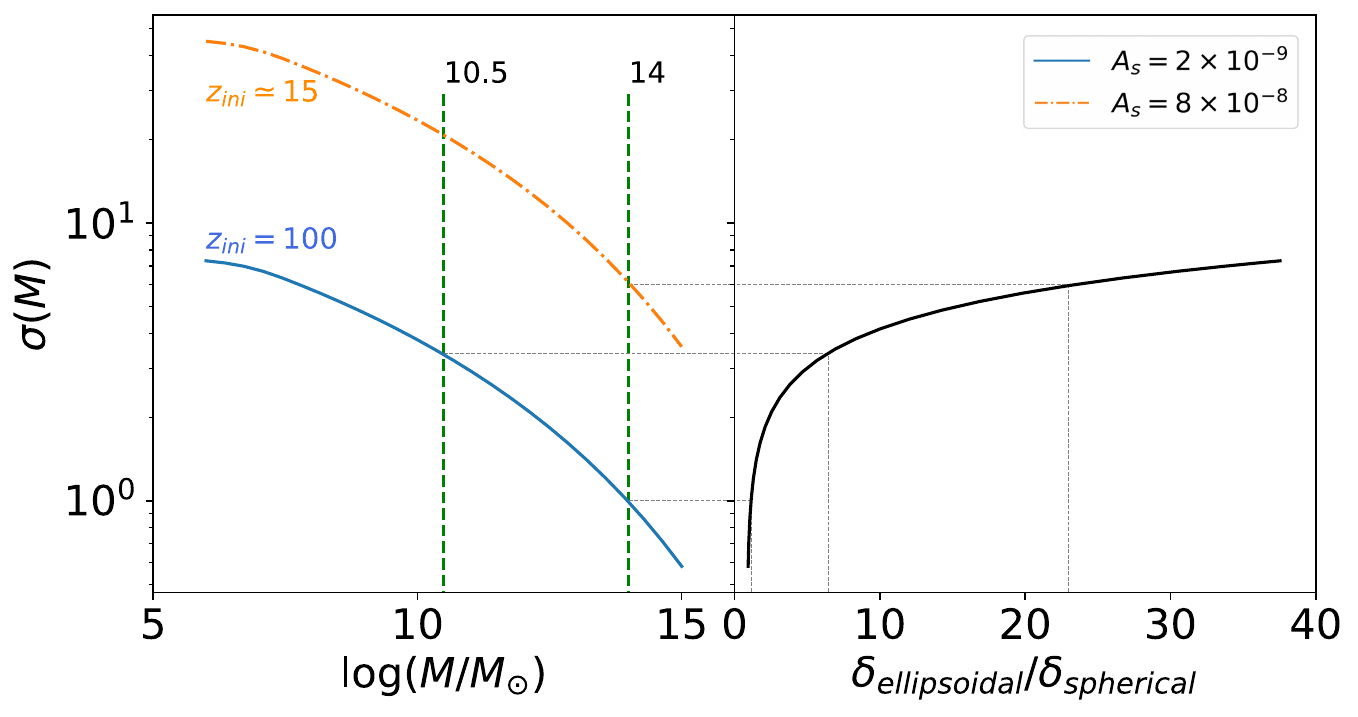}
\caption{The variance of linear matter perturbations is plotted versus mass (left panel) and versus the ratio of the critical density of ellipsoidal to spherical (right panel). In the left panel, the variance is plotted for two different initial power spectrum amplitudes $A_s=2\times 10^{-9}$ corresponding to initial redshift $z_{ini}=100$, and $A_s=8\times 10^{-8}$ assigned to $z_{ini}=15$. The green dotted vertical lines show the scope of the mass range under study.} 
\label{fig:1234}
\label{fig:theory}
\end{figure}
To conclude this section, we refer to Fig.\ref{fig:1234}, in which we plot the variance of perturbation in terms of DM halo mass in the left panel, for two amplitudes of the initial power spectrum. The vertical dashed green lines show the mass range in our simulation. In the right panel, we plot the variance (same y-axes for both panels) in terms of the ratio of the critical density of ellipsoidal to the spherical model introduced in equation \ref{eq:ellip-barrier}.
Note that the variance is a monotonic decreasing function of mass, and the critical density ratio is a monotonic increasing function of mass. This is a crucial point to assert that by increasing the amplitude of perturbations for a specific mass range (orange dash-dotted line) we obtain higher variances. It means that by fixing the variance, a mass range in the cosmology with higher initial amplitude corresponds to a smaller mass scale in the $\Lambda\text{CDM}$ paradigm, where the velocity shear field is crucial in the collapse model. \citep{Sheth:2001dp, Kameli:2019bki}.
{{The equivalent interpretation is that the higher amplitude mimics an initial box at a later time (lower redshifts). In Fig.\ref{fig:1234}, we show the corresponding initial redshift. The $A_s=2\times 10^{-9}$ correspond to initial redshift $z_{\rm ini}=100$ and $A_s=8\times 10^{-8}$ correspond to $z_{\rm ini}\simeq 15$ . Accordingly, due to the right panel, we are probing the scales in which the ratio of critical densities is higher than unity. The unity ratio of critical densities means we have almost spherically collapsed halos. However, a large ratio indicates that we should consider ellipsoidal collapse as velocities play an important role. Another way to put this is that in later times, the system becomes non-linear, and information on velocities is needed. It is a key idea in this work, which we will discuss further in upcoming sections. }}
\section{$N$-body Simulations}
\label{sec3}
\begin{figure}
\centering
\includegraphics[width=\linewidth]{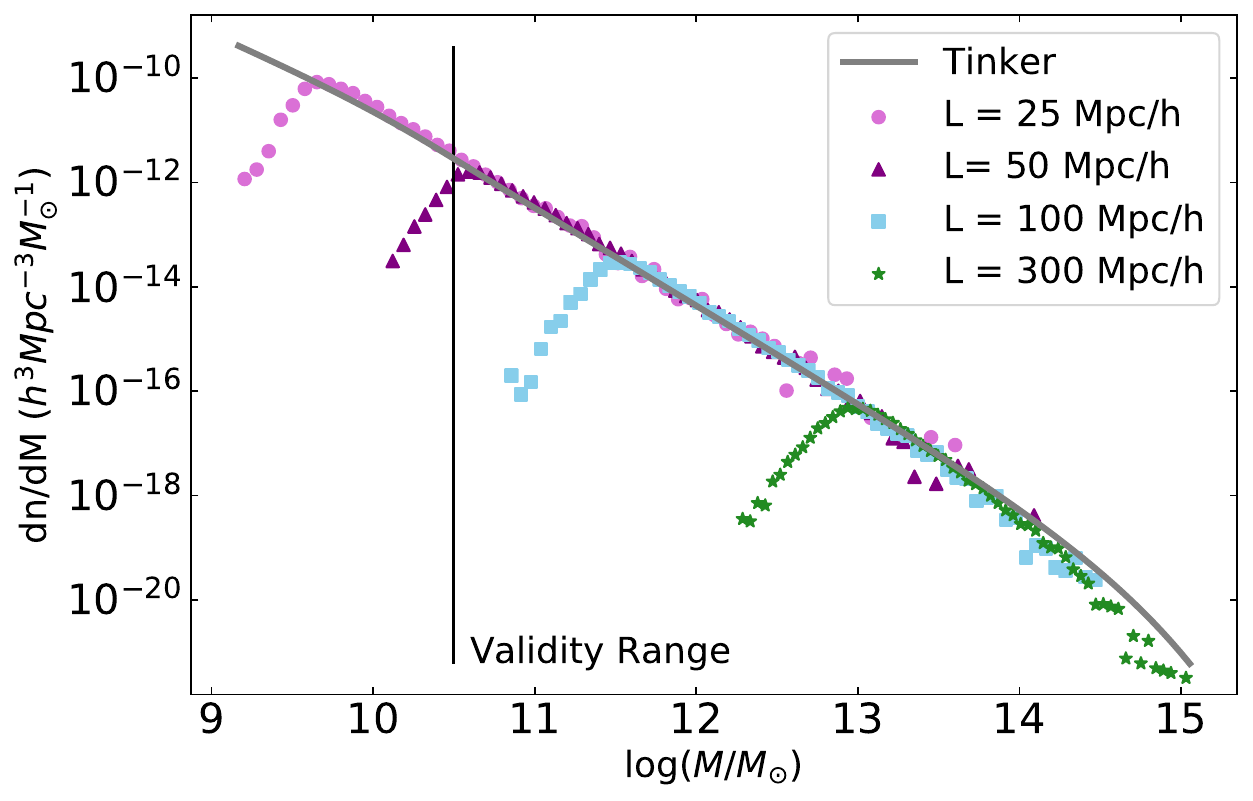}
\caption{Halo mass function for simulation boxes with different sizes and theoretical halo mass function from the Tinker profile is plotted. For box size $50 {\text{ Mpc}}/h$ we have the validity range for halo masses up to $2\%$ in comparison to the Tinker profile. This is for the mass range of $10^{10.5}M_{\odot}$ to $10^{14}M_{\odot}$. For all the box sizes, the number of simulating grids is 600, so the number of dark matter particles in the box will be $600^3$. }
\label{hmf_sim}
\end{figure}
To study the structure formation using deep learning methods, we use data obtained from gevolution, a relativistic cosmological $N$-body simulation \citep{Adamek:2015eda,Adamek:2016zes}.
In gevolution, the full set of Einstein's equations is solved to update the positions and velocities of particles. 
One of the main advantages of gevolution is its natural extendability to consider relativistic particles (e.g., massive neutrinos \cite{Adamek:2017uiq}) and non-standard cosmologies (e.g., different dark energy models and modified gravity theories) \cite{Hassani:2019lmy, Hassani:2019wed, Hassani:2020rxd}. 
The presence of a horizon in relativistic equations compared with the Newtonian ones provides a more accurate analysis of the structure formation at large scales. The gevolution's potential for covering the relativistic particles and seeing their effects in the matter power spectrum on small scales as well as large scales shows us a wide range of exciting possibilities to investigate the effects of non-standard cosmological models (like assuming massive neutrinos or considering different dark energy models) on the perturbation theory and formation of structures in our future works.\\
Assuming the standard model of cosmology, $\Lambda$CDM, we run our simulations with a standard cosmological model configuration. Our main training data consists of 10 gevolution boxes with 50 ($\text{Mpc } h^{-1}$) box length and $600^3$ DM particles within the box. We use the cosmological parameters $\Omega_{\text{cdm}} h^2 = 0.122$, $\Omega_{\text{b}} h^2 = 0.021$, $T_{\text{CMB}} = 2.725$, $ h= 0.67$ , $A_s = 2\times 10^{-9}$, $n_s =0.96$ and $N_{\text{eff}} = 3.046$. We made use of the public halo-finder code Rockstar \citep{Behroozi:2011ju} to identify halos at $z=0$ in the box.\\ 
Due to the fixed mesh and finite resolution in our simulation data sets, there is an inaccuracy in finding small-mass halos.  Thus, to find the best mapping between the initial condition of dark matter particles and their final halo mass, we should recognize the resolution effect on the halo mass function. The resolution limitation is not physical and is only due to the simulation or halo-finder imprecisions. So we have to choose our training data-set from the particles that end up in the halos with proper mass in our specific training box. To find the resolution effect, we run four simulations with different resolutions, where the number of particles is fixed to $600^3$ but different box sizes (25, 50, 100, and 300 $\text{Mpc } h^{-1}$) are considered. The overlapped region of the halo mass function for the different resolutions shows the validity range of each simulation. In Fig.\ref{hmf_sim}, we show the validity range for each simulation. Based on the figure the validity range for the box size 50 $\text{ Mpc } h^{-1}$, which is the main simulation being used in this study, within $2\%$ accuracy is $ 10^{10.5} M_{\odot}$ to $ 10^{14} M_{\odot}$. \\
\section{Method: Convolutional Neural network}\label{sec4}
In this section, we discuss the ML methods we use in this work. In the first subsection \ref{sec4.1intro}, the convolutional neural network is discussed. In the upcoming subsections, the image construction \ref{sec4.2image}, architecture \ref{sec4.3arch}, loss function \ref{sec4.4loss}, image size \ref{sec4.5size} and image resolution \ref{sec4.6res} are described respectively.
\begin{figure*}
\centering
\includegraphics[scale=.5]{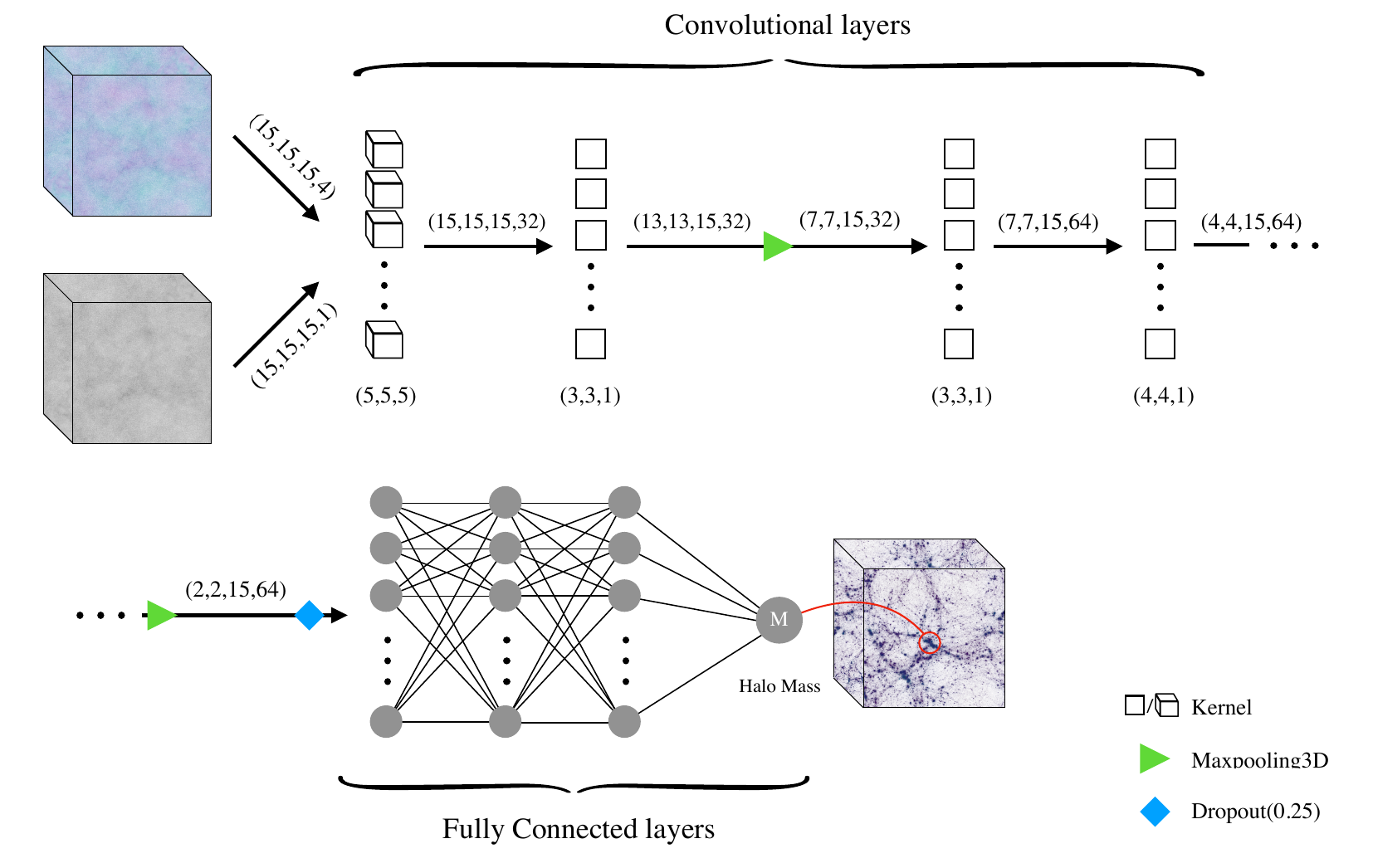}
\caption{{We train our CNN on a set of cosmological simulations with different phase space information. Cosmological simulations alone cannot provide us with insight into the physical processes in structure formation and its relation to the main parameters in the initial condition, which are important in halo mass function. We use our neural net, which is trained on different sets of initial condition information of the simulation box to provide an understanding of the structure formation. We demonstrated our two kinds of training in this plot. We once train our model on the images with the velocity field information ($15^3 \times 4$ input shape) and once without($15^3 \times 1$ input shape) to investigate the effect of the velocity field in our machinery's learning.}
}
\label{fig:model}
\end{figure*}
\subsection{Convolutional Neural Networks}
\label{sec4.1intro}
Convolutional neural networks (CNNs) \citep{10.5555/2969644.2969648} are one of the most powerful types of neural network models and are widely used for image processing applications. Inspired by Neocognitrons \citep{Fukushima1980}, they benefit from locality-based operations to hierarchically detect translation-invariant patterns and make predictions based on them. Also, they use parameter sharing, which reduces their complexity compared to traditional neural networks. This allows for building extremely deep neural networks that can learn complex patterns efficiently. From the early applications of CNNs in \cite{6795724} and \cite{726791} to the modern applications in life sciences \citep{rooster_book}, medical imaging \citep{LUNDERVOLD2019102, Yamashita2018} and autonomous driving \citep{Grigorescu2020}. CNN models have always been one of the driving forces in the field of deep learning.

In recent years, applications of CNNs in cosmology have attracted a lot of attention. This is partly motivated by the big data from cosmological observations and large-scale simulations \citep{siemiginowska2019astro2020}. As a more specific example of data-driven cosmology, we follow the line of the research in the LSS, using $N$-body simulations as an arena to study the distribution of matter.\\
For instance, \cite{Ravanbakhsh:2017bbi} used a CNN to estimate cosmological parameters directly from zero-redshift density field images and outperformed conventional methods. In \cite{Zhang:2019ryt}, U-net architecture has been used to infer baryonic matter distribution from less costly dark matter only $N$-body simulations. \cite{Villaescusa-Navarro:2021pkb} used a CNN model to learn moments of marginalized joint posterior of cosmological and astrophysical parameters, and thereby performed likelihood-free inference to work out $\Omega_m$ and $\sigma_8$ from 2D field maps. 

Like other deep learning models, CNN-learned feature maps are not easy to interpret. Thus, deriving theoretical conclusions from such models is challenging. Some works have been dedicated to developing methods to interpret neural networks \citep{ribeiro2016why, Mahendran_2016, MONTAVON20181} or to implement easily interpretable networks \citep{zhang2018interpretable}. 
It is also possible to shed some light on deep neural networks by training them using dissimilar or feature-reduced data sets and comparing the results. For example, \cite{Lucie-Smith:2020ris} tried to interpret the learned feature maps of a CNN to assess the importance of anisotropic features in the prediction of dark matter halo masses. We employed the latter method to conclude our CNN's trained parameters.

We develop a deep CNN to learn about the physical processes of structure formation directly from $N$-body simulations. We train our CNN to learn the halo mass that each particle resides in at the redshift zero. It is worth mentioning that the halo mass is defined based on the number of bound particles, thus there is no assumption on the shape of halos in our analysis. Our CNN's input is the density and velocity field information at $z=100$. We use our trained network to investigate the effect of the velocity field in halo collapse and the final halo mass function. In the upcoming subsections, we first describe our data preparation and image construction and then give an overview of the architecture of our CNN model.
\subsection{Image Construction}
\label{sec4.2image}
We use one of the most basic features of the simulation box, namely the phase space information of the DM particles.
We make a 3D cubic image of the local neighbourhood of each particle at the initial snapshot of simulation boxes at $z = 100$. Each image consists of the particle itself at the central voxel. We use $0.5 \text{ Mpc }h^{-1}$ comoving size for resolution and an image size of $15$ voxels corresponding to $7.5 \text{ Mpc } h^{-1}$ comoving size for our model.  We prepare the density and velocity field from the phase space by averaging the information of the particles for each voxel.
Images consist of four channels, one for the normalized number of particles in each image pixel and three for the net velocity of particles in the three Cartesian coordinates. We want our model to predict the mass of each particle's final halo destination at the last snapshot $z=0$. For this task, first, we should fix the image resolution and the image size. The combination of $7.5 \text{ Mpc } h^{-1}$ image size and  $0.5\text{ Mpc } h^{-1}$ resolution is the conservative image properties that give us a reasonable run time while preserving the essential information. We will study in more detail the choice of these two parameters in the next section after describing our deep learning model and method. 
The input of our ML model is the image constructed for each particle, and the output is the particle's final destination halo mass. 

In order to decrease our machine's sensitivity to different simulation realizations and to reduce the effect of Poisson error, we use 10 independent simulations. Since the number of large halos in the simulation dramatically decreases due to small box size and mass resolution consideration, our model may face a sample variance problem as it can not see enough halos at a large mass range. Considering the independent simulations decreases the Poisson error due to the small number of large halos by a factor of one. 
To reduce it by a larger factor of two, we need 100 independent simulation boxes, which is challenging based on our computational power. 
We pick 300,000 particles randomly from all these boxes as our training sample. Increasing this number from 300,000 to 400,000 showed minor improvement in the result while adding to the computational cost extensively.

We make the images for each particle in the way described above. We exclude all the particles ending into a halo with masses less than the allowed halo mass range by simulations in both training and testing sets. This choice of minimum halo mass is compatible with the main goal of this work, namely, studying the effect of the velocity field information. Note that the particles ending in cosmic web structures other than halos such as voids and filaments are not considered in this study; however, they are present in the image of the particles. One will need a simulation with a larger resolution to broaden the allowed halo mass range for a comprehensive study of structure formation.
\\
Based on the physical properties of structure formation, we expect the inferred halo masses to be invariant under reflection and rotation of the input subregions. In order to meet such symmetries, we use 48 members of the cube symmetry group to replicate our input images. This has been shown to effectively reduce the variance of predictions within the replicates of each subregion in \cite{Ravanbakhsh:2017bbi}. However, in order to speed up the training process, we do not use the whole replicates to augment our data-set. Instead, the data set the generation of each mini-batch, each input image is transformed by a randomly selected member of the symmetry group. While preserving the possibility of feeding our model with different transformations of each subregion, this approach considerably reduces our training time. 
\subsection{Architecture}
\label{sec4.3arch}
\begin{table}
\begin{center}
\resizebox{\linewidth}{!}{
\begin{tabular}{ l c | c | c | c | c | c | c |}
\cline{1-7}
&&&&&&\\
\multicolumn{2}{|c|}{\textbf{Input:}} & \multicolumn{3}{|c|}{Subregion around a particle}  & \textbf{shape:}  & (15, 15, 15, 4) \\
&&&&&&\\
\cline{1-7}
layer & type & kernel size & padding & activation & param \#  & output shape\\
\cline{1-7}
1 & Conv3D & (5, 5, 5) & same & sigmoid & 16032   & (15, 15, 15, 32)\\
\cline{1-7}
2 & Conv3D & (3, 3, 1) & valid & relu & 9248   & (13, 13, 15, 32)\\
\cline{1-7}
3 & Maxpooling3D & (2, 2, 1) & same & - & 0   & (7, 7, 15, 32)\\
\cline{1-7}
4 & Conv3D & (3, 3, 1) & same & relu & 18496   & (7, 7, 15, 64)\\
\cline{1-7}
5 & Conv3D & (4, 4, 1) & valid & relu & 65600   & (4, 4, 15, 64)\\
\cline{1-7}
6 & Maxpooling3D & (2, 2, 1) & valid & - & 0   & (2, 2, 15, 64)\\
\cline{1-7}
7 & Dropout (0.25) & - & - & - & 0   & (2, 2, 15, 64)\\
\cline{1-7}
8 & Flatten & - & - & - & 0   & (3840)\\
\cline{1-7}
9 & Dense & - & - & relu & 460920   & (120)\\
\cline{1-7}
10 & Dense & - & - & relu & 7260   & (60)\\
\cline{1-7}
11 & Dense & - & - & linear & 61   & (1)\\
\cline{1-7}
&&&&&&\\
\multicolumn{2}{|c|}{\textbf{ Output:}} & \multicolumn{3}{|c|}{ Destination halo mass }  & \textbf{shape:}  & (1) \\
&&&&&&\\
\cline{1-7}
\end{tabular}}
\end{center}
\caption{{The architecture of the sequential Convolutional Neural Network model.}}
\label{table:CNN}
\end{table}
Our model is a deep convolutional neural network which its architecture is described in Table \ref{table:CNN}. The model has two parts: a contracting path and a fully connected component. The contracting path consists of two scales. At each scale, the input is passed through two convolutional layers, followed by a max pooling layer. In each convolutional layer, a set of filters are convolved with the input and the result is passed through an activation function. These operations are used to extract features from the input. Max pooling layers down-sample the feature maps on each scale while preserving their important structural information. This is done by sliding a kernel across each image channel and putting the maximum value of each resulting region in the corresponding location of the respective output channel. Fully connected layers use the feature maps extracted by the contracting path to infer the scalar value of the halo mass to which the central particle of the input image will finally belong (see Fig.\ref{fig:model}).
\subsection{Loss Function}
\label{sec4.4loss}
\begin{figure*}
\centering
\includegraphics[scale=.35]{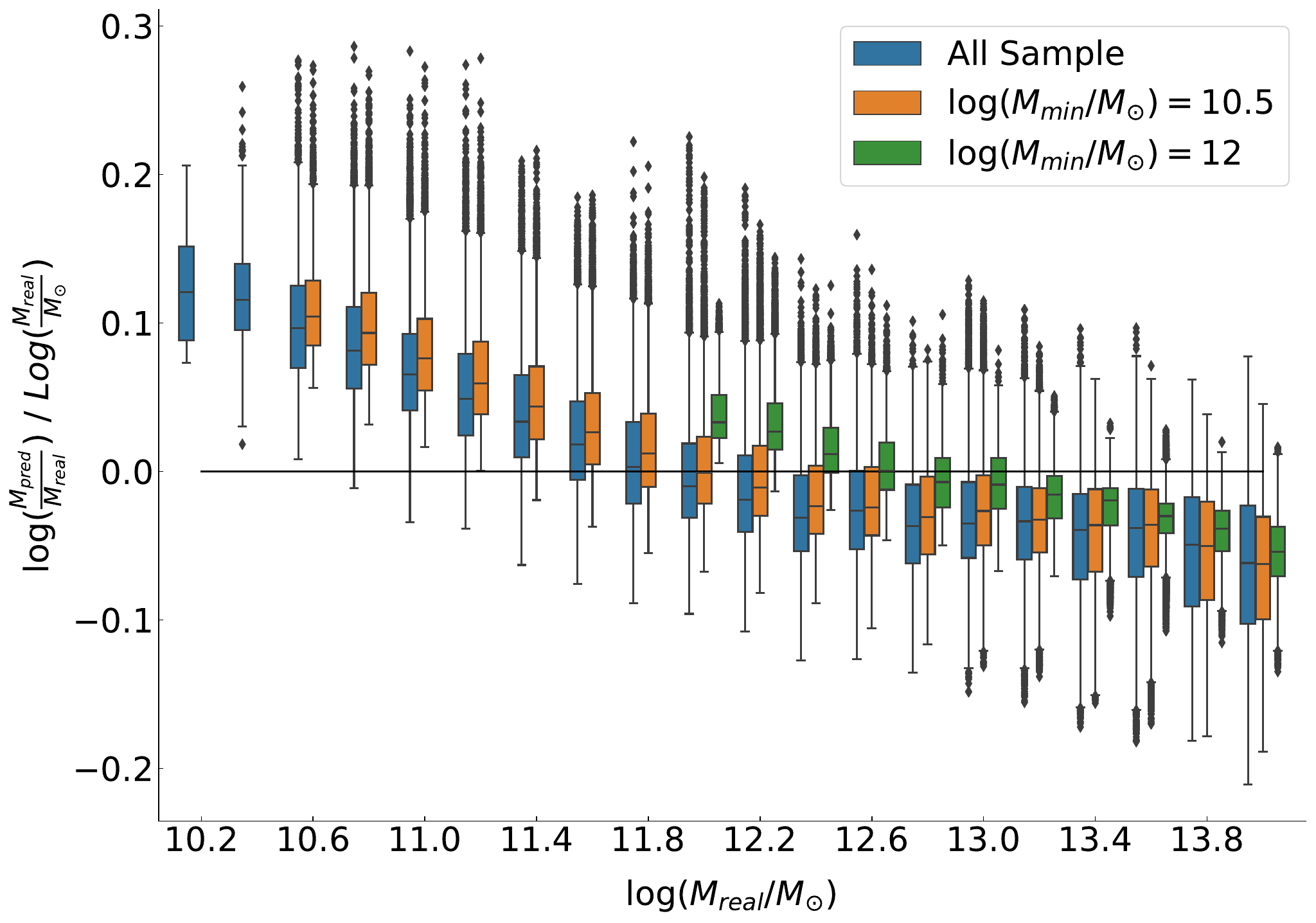}
\caption{The mass of the predicted dark matter halos to the real mass sample is plotted for three minimum mass ranges of the machine entry data. The error bars show the mean and the $\pm 25\%$ of the distribution. 
Different colours show the minimum mass which is chosen for ML. The presence of the non-symmetric outliers shows the systematic overprediction of our model at the smallest halo ranges. We resolve this issue by keeping the complete halo mass range of the box while only trusting the results on an interval. This technique pushes the over-prediction to the mass ranges, which are out of the range of interest.}
\label{min_mass}
\end{figure*}

The objective loss function used for the model is mean squared error (MSE) which is minimized using the Adam optimization algorithm \citep{kingma2017adam}. As we will demonstrate later, our model with this symmetric loss function introduces a systematic error, which causes us to over-estimate smaller mass ranges. \\
In Fig.\ref{min_mass}, we see an over-prediction for the mass of smaller halos. This feature is present in all samples, indicating an intrinsic error in the ML method. Accordingly,  as the over-prediction increased with a higher minimum mass range,  we will not restrict our sample with this consideration. Even more, we use the smaller mass range which is not in the range of our model prediction. This is done in order to transfer the over-prediction error to a smaller mass range which is out of the scope of the study.
As demonstrated in Fig.\ref{min_mass}, we can see that changing the minimum halo mass given to the CNN model does not affect the prediction of the model more than the systematic error due to the symmetric loss function that we discussed above. This means that predictions of the halo mass are not affected by slightly changing the minimum used in our training and thus it is independent of our choice of halo mass range in the training set.  

\subsection{Image size}
\label{sec4.5size}
\begin{figure}
\centering
\includegraphics[width=\linewidth]{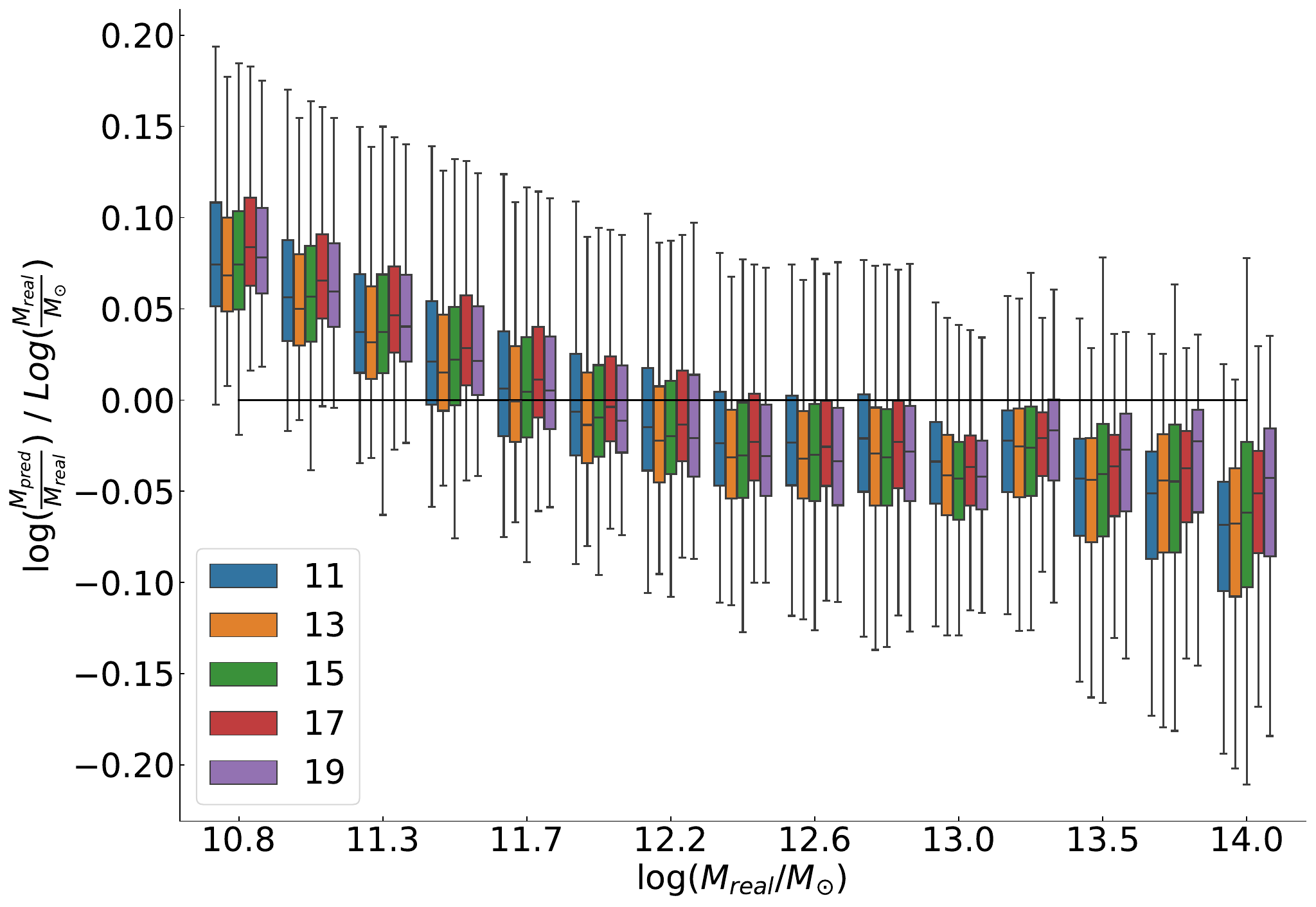}
\caption{The mass of predicted dark matter halo concerning real one is plotted versus mass for different image sizes. The different colours correspond to different image sizes ina units of $\text{Mpc } h^{-1}$. The error bars show the $\pm 25\%$ of the prediction distribution in each mass bin. }
\label{size-fix}
\end{figure}
One of the critical hyper parameters of our model is the physical size of the input images. The initial density contrast field at $z_{ini}=100$ is almost homogeneous and less than unity. Furthermore, for a desired final halo, we can approximate the initial protohalo (ph) size as $V_{\text{ph}} = M_{\text{halo}} / \bar\rho_{i}$, where $\bar\rho_{i}$ is the average density of the simulation box at $z_{ini}=100$. As a result, the image size should be at least the volume needed to contain all the most massive halo particles. 
There is a trade-off in choosing our image size.  Our mapping between these images to the halo mass only aims to find the future host halo's mass of the central part of the image. Besides their substantial computational cost, large-volume images may mislead the machine when we are not on the density field peaks, and many proto-halo regions exist within a single image.
Also, we can not choose our images too small, as we miss some information necessary to find the future halo mass of the specified region in the initial box. We do some scale tests for choosing the best possible image size. We change the image size slightly around the value we obtained as described and compare our model's performance with different image sizes. Fig.\ref{size-fix} indicates that the error due to image size is almost the same for all sizes that we investigate. 
In general, we can categories the systematic errors involved in our machinery as a) error due to the resolution of the simulation (see Fig.\ref{hmf_sim}), \\b) the systematic over-prediction error (see figure \ref{min_mass}), and \\c) error due to the characteristic of the image (Figs.\ref{size-fix},\ref{res_fix}).\\
We set an upper limit of $20\%$ for our uncertainty, and for the rest of the work, we choose a specific set of hyper parameters for our images concerning the three sources of error mentioned earlier. This means the minimum mass is set to the minimum halo mass in simulations as described in section \ref{sec4.4loss}, and the image size is set to $7.5 \text{ Mpc }h^{-1}$ ($15$ voxels).
\\
\subsection{Image resolution}
\label{sec4.6res}
\begin{figure}
\centering
\includegraphics[scale=.25]{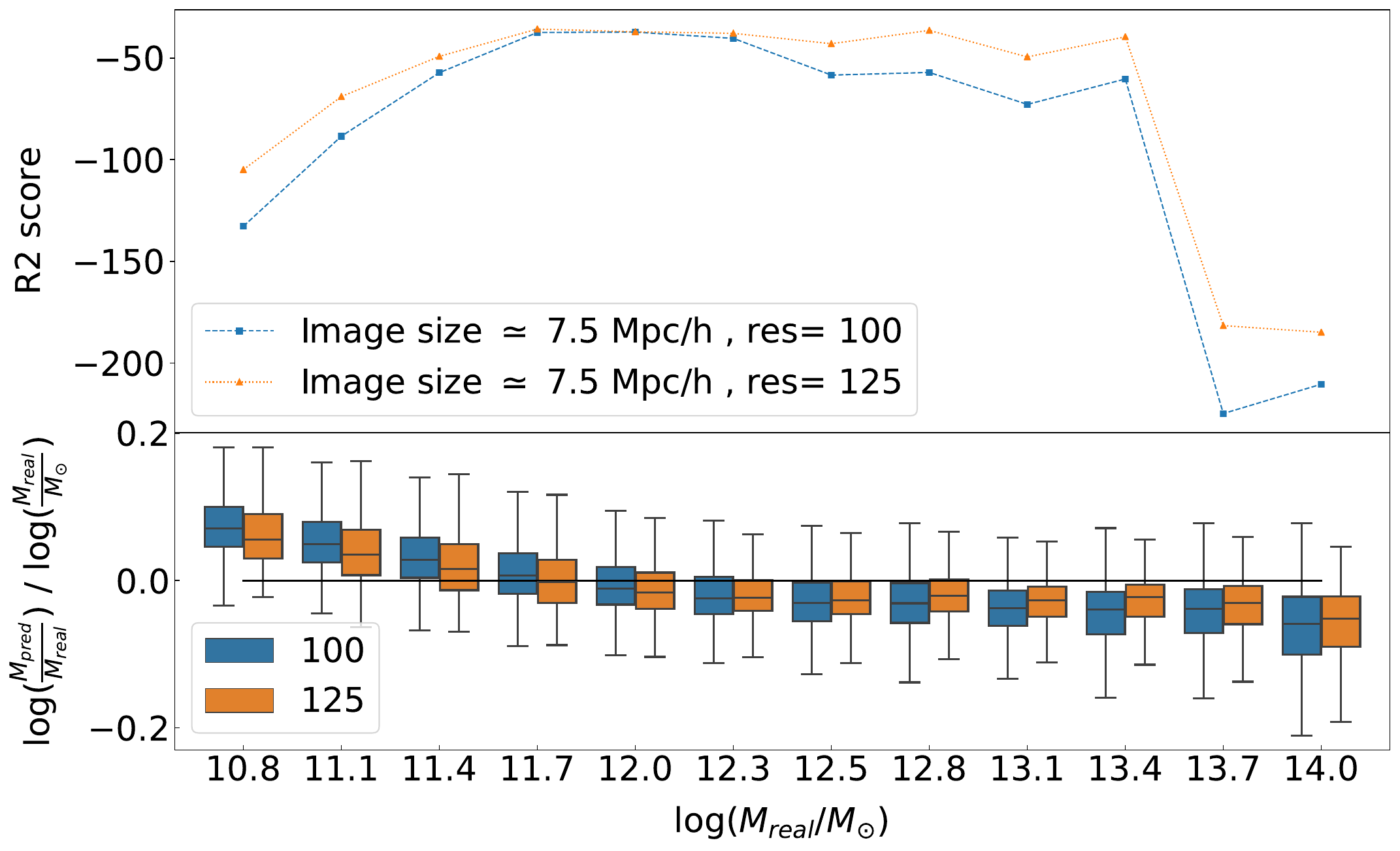}
\caption{ The relative error and R2-score of the predicted final halo mass with respect to the real one is plotted versus mass for different image resolution of $100^3$ and $125^3$ voxels for a $50 \text{ Mpc } h^{-1}$ box. }
\label{res_fix}
\end{figure}

After fixing the appropriate image size, we specify the image resolution for the CNN input. The volume of the minimum halo mass in the initial condition is the lower bound for the resolution. Thus we pick a value using this minimum bound, and then we change our model's resolution around this specific value to reach the best resolution. As we show in Fig.\ref{res_fix}, increasing the image's resolution by a factor of $1.25$ does not change the overall accuracy of the model in the desired mass ranges noticeably. Note that this is done while keeping the physical length of the image constant. However, this increase in the resolution adds to the computational cost by more than a factor of two. Furthermore, resolution $100$ (physical voxel size of $0.5 \text{ Mpc } h^{-1}$) is a convenient choice.
\section{Results}\label{sec5}
We use the basic phase space information of the initial simulation box with a finite resolution smoothed with a cubic window function, as the input of our CNN model. We make images of the particle's local neighbourhood and train our machine to predict the mass of the final halo to which this particle ends up from its local neighbourhood picture. We use a training set of $300,000$ images randomly chosen from 10 different realisations of the same cosmological parameters. Using the results described in section \ref{sec4} to fix the hyper-parameters of the CNN model, we set the image size to be $7.5 \text{ Mpc }h^{-1}$ or $15$ pixels and with a resolution of $0.5 \text{ Mpc } h^{-1}$. 
Our results can be separated into two main parts. First, we use our model to recover the halo mass function of a completely new box but with the same cosmological parameters as the training boxes in subsection  \ref{sec hmf}. Then we present a physical interpretation of the structure formation using our framework based on the effect of the initial velocity field in the prediction of HMF in subsection \ref{sec velocity}.  
\subsection{Halo Mass Function Prediction}
\label{sec hmf}
\begin{figure}
\centering
\includegraphics[scale=.5]{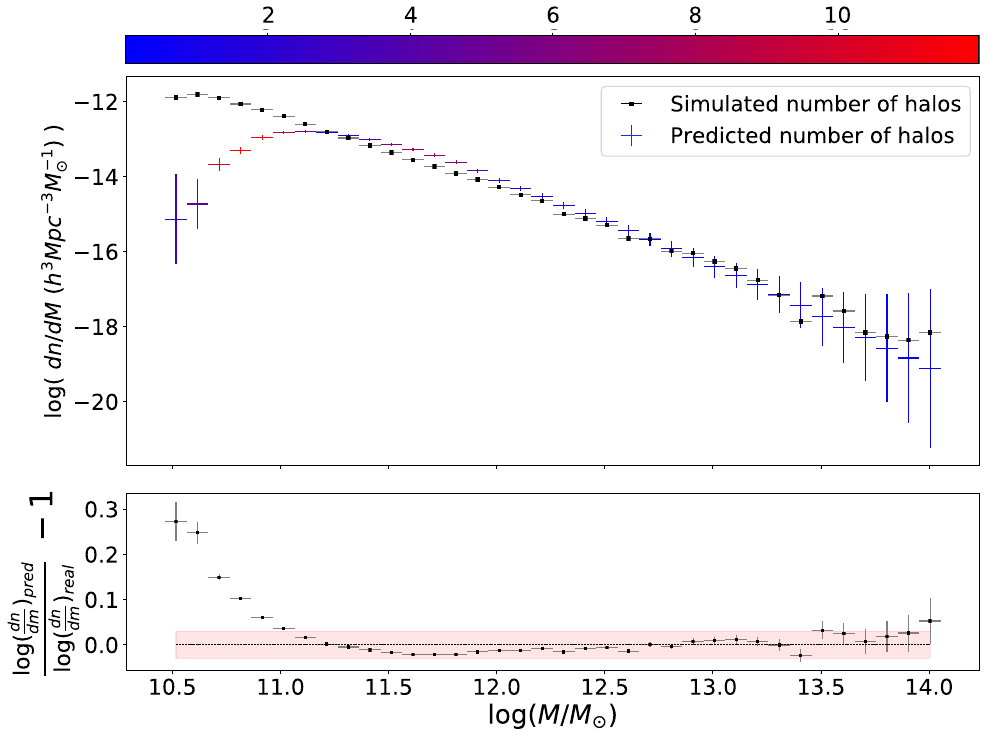}
\caption{The number density of predicted and real dark matter halos plotted versus mass in the upper panel. The colour bar shows the $\sigma$-deviation of our predicted model compared to the real data. In the lower panel, we show the ratio of the number density, and the shaded region shows the $3\%$ confidence level. In both panels, we show the model which considers the velocity as well as density field as an input.}
\label{hmf}
\end{figure}

The aim of constructing an accurate predictor for halo masses is to eventually predict the halo mass function, given an initial distribution of particles. Furthermore, we use the distance of our predicted halo mass function from the real halo mass function as a metric to measure the performance of our model. To predict the halo mass function, we first need to apply a binning on our regression problem and then find the predicted number of halos corresponding to each halo mass bin. As the prediction the of halo mass function is strongly dependent on the bin number, we should be careful about our binning definition. We define 40 bins between $10.5 \leq \log(M/M_{\odot}) \leq 14$ linear in the logarithm of mass. By applying the trained model on a set of $280,000$ randomly chosen particles in our test box and predicting the halo mass for each of them, we already acquired the number of particles in each halo mass bin. Next, we use the average density of each halo mass bin to convert the number of particles to the number of halos in that bin. One also must scale this result by the relative ratio of the total number of randomly chosen particles used in this prediction over the total number of particles in the box.
The main two contributing factors in the error of this distribution are first, the systematic error in the prediction of halo masses arising from our CNN model's imprecision, and the second error comes from the lack of enough data in chosen random point.  
Each error is shown in Fig.\ref{hmf}, where we demonstrated the predicted halo mass function of the box. At our ending mass bins, where the sample variance due to the lack of most massive halos increases, our confidence in prediction drops. The colour bar in this plot corresponds to the statistical error with an n-sigma deviation of our prediction from the real HMF of the box which is our metric. Our prediction for the HMF is in agreement with the real one from the simulations in the most mass ranges. However, our prediction ability drops considerably for the mass ranges of $\log(M/M_{\odot}) < 11$, which is because of the effect we described in section \ref{sec4}. This result shows that our model prediction is trustable the most in the range of  $11< \log(M/ M_{\odot}) < 13$. 

\subsection{Information content of the velocity and density fields}
\label{sec velocity}

{{We aim to understand whether our trained CNN model can excavate the most relevant information in the boxes' initial velocity and position fields for the reconstruction of the late-time halo mass function. In the linear regime, the natural expectation is that due to the decay of the growing modes in the density perturbations, the relevant dominating modes are those completely solvable using the Poisson equation. Thus, the linear regime's initial velocity and density fields must have the same information content. To check the fit of our CNN model, we first check to see if our CNN is also finding the velocity and density field as equivalent to each other in the linear regime. In this first case, we use our CNN to reconstruct the late-time halo mass function at $z=0$ from the initial information of a cosmological box in the $z_{ini}=100$ with an amplitude of perturbations equal to the standard value $A_s = 2\times 10^{-9}$, and the mass range of interest provided in the previous sections.
 
To investigate the effect of the velocity field in the prediction of our CNN maps, we train our model on different sets of input features. One only consists of the density field information, while the other one has information of velocities averaged over a certain scale. }} We change our CNN model in each case to fit the input image shape ( 15,15,15,1) for the case without velocity information and (15,15,15,4)  for when the velocity information is also provided. Except for the first layer, which is changed to fit the input's shape, all the other layer's architecture remains the same.
\begin{figure}
\centering
\includegraphics[scale=.25]{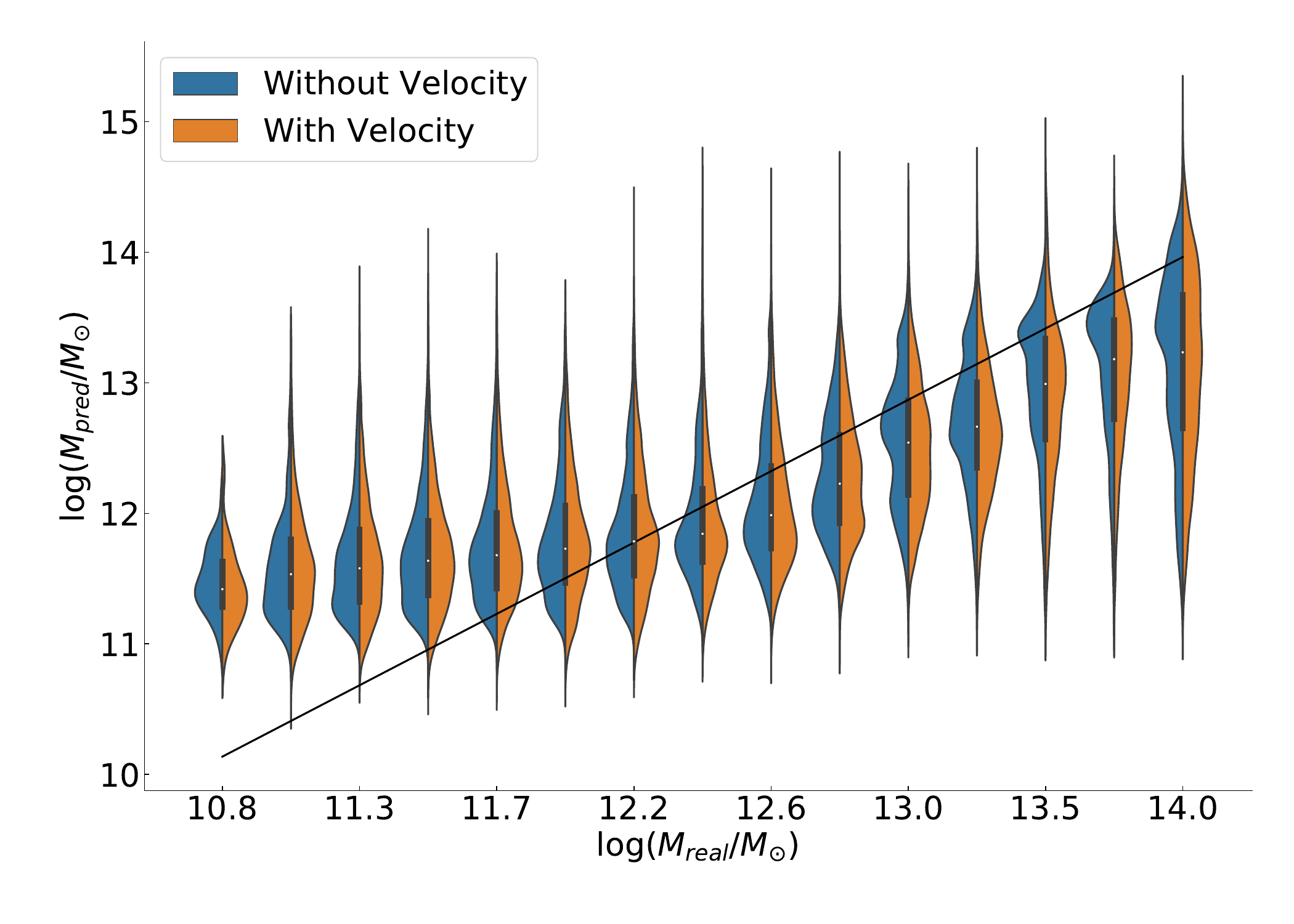}
\caption{The predicted mass of dark matter halos versus the real mass is plotted. The violins show the distribution of our prediction for two different sets of training data with velocity (blue/left) information and the other without velocity (orange/right) information.}
\label{violin}
\end{figure}
In Fig.\ref{violin}, we show the prediction of halo mass with and without considering the velocity information. 
Violins represent two different sources of error in each halo mass bin. One is the distance of the solid line to the distribution average for the predicted halo mass in each mass bin (the accuracy of the result), and the other is the variance of the distribution for the predicted halo mass (precision of our prediction). To consider these two error sources, we use the R2-score, which would be a reasonable metric to see whether there is an improvement in the prediction with and without velocity in each halo mass bin.\\
The advantage of this metric is that by dividing the model's standard deviation, it measures the score relative to the model itself and thus enables us to compare the score of the two different models.
We use R2-score measurement to quantify the performance of our model, where the R2-score is measured in each mass scale as:
\\
\begin{align*}
    \text{R2-score} (y_{\text{real}} , y_{\text{pred}}) = 1 - 
    \frac
    {\sum_i (y_\text{real}^{(i)} - y_\text{pred}^{(i)} )^2}
    {\sum_i (y_\text{real}^{(i)} - \bar{y})^2} \ ,
\end{align*}
\\
where $y_\text{real}^{(i)}$ is the real mass of the particle's destination halo and $y_\text{pred}^{(i)}$ is the prediction of our model for it, and the
 $\bar{y}$ is the average $y_{\text{pred}}$ on that mass bin.
We measure the R2-score on different mass scales for the two configurations. One with and the other without velocity information in the case of standard initial power amplitude 
$A_s = 2\times 10^{-9}$. {{Note that the overall R2-score for the models would be a value between 0 and 1, as expected from its definition, while when we measure the R2-score on individual bins, it can get other values as well. Being negative on a bin would only mean we use small bin sizes compared to the complete mass range.
We use the cross-validation technique to find the error bars on our model. In each case, we train our model five separate times on randomly chosen training sets of the same size. We measure the R2-score for each of these different trained models at each mass scale and use the mean and standard deviation of the bootstrapped sample as our model's mean and uncertainty in terms of the R2-score.
\begin{figure}
\centering
\includegraphics[scale=.4]{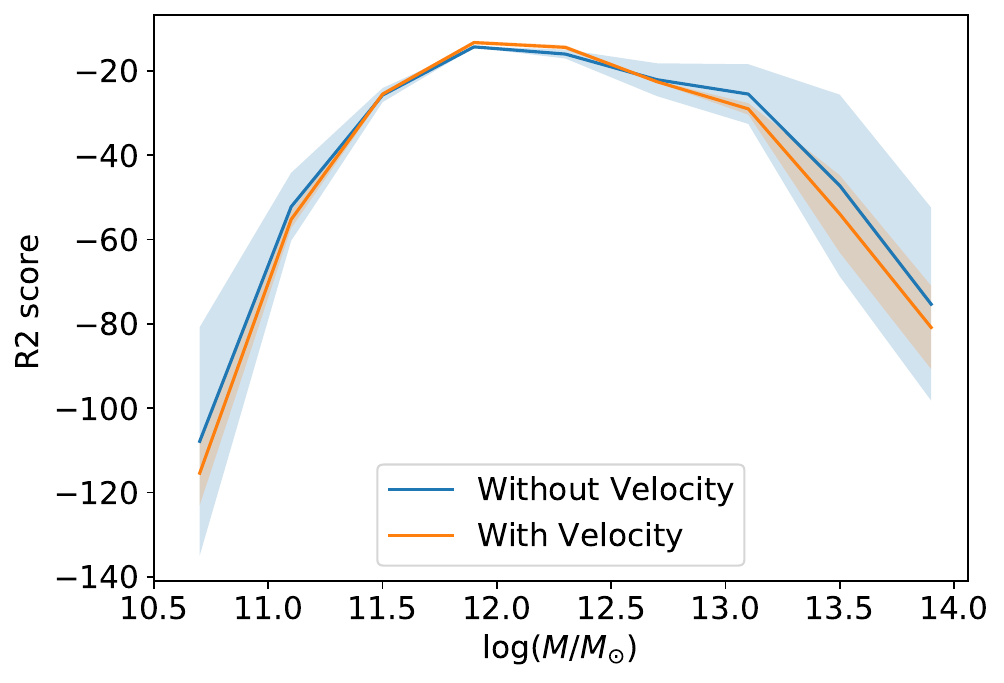}
\caption{ R2-score for the two with and without velocity models. The result shows that there is a negligible difference between the performance of these two models in the standard model of cosmology with initial curvature perturbations of $A_s = 2\times10^{-9}$. }
\label{cross-val}
\end{figure}
Based on Fig.\ref{cross-val}, the mean R2-score value does not change considerably for the model without velocity in comparison with the one with velocity field information. This result in Fig.\ref{cross-val} demonstrates that in the mass range $10.5 \lesssim \log(M/M_{\odot}) \lesssim 14$, our CNN maps find sufficiently equivalent information in the velocity and density fields that is to be expected. Given this result, our CNN map successfully passes our sanity check, providing that this map can solve the Poisson equation. \\
However, due to our discussion in section \ref{sec2}, we anticipate that the velocities should be essential in small mass scales or later times, where the non-linear effects are no longer negligible. Nonetheless, due to the mass resolution limitations of our simulations, this regime is not accessible to our current computational power and configuration.
To overcome this obstacle, we use an idea to check the velocity dependence of the structure formation. We increase the initial power spectrum of the perturbations in the $N$-body simulation runs. It causes dark matter halos in our simulation to mimic the velocity field in smaller mass range halos due to their higher over-density, as discussed in Sec.\ref{sec2}.
Moreover, as discussed in Sec.\ref{sec2}, the increase of the amplitude can also mimic an effect of decreasing the redshift and moving to the later times where non-linear effects appear in our chosen mass range. It means that we are investigating the PS idea with lower initial redshifts. }}

\begin{figure}
\centering
\includegraphics[scale=0.45]{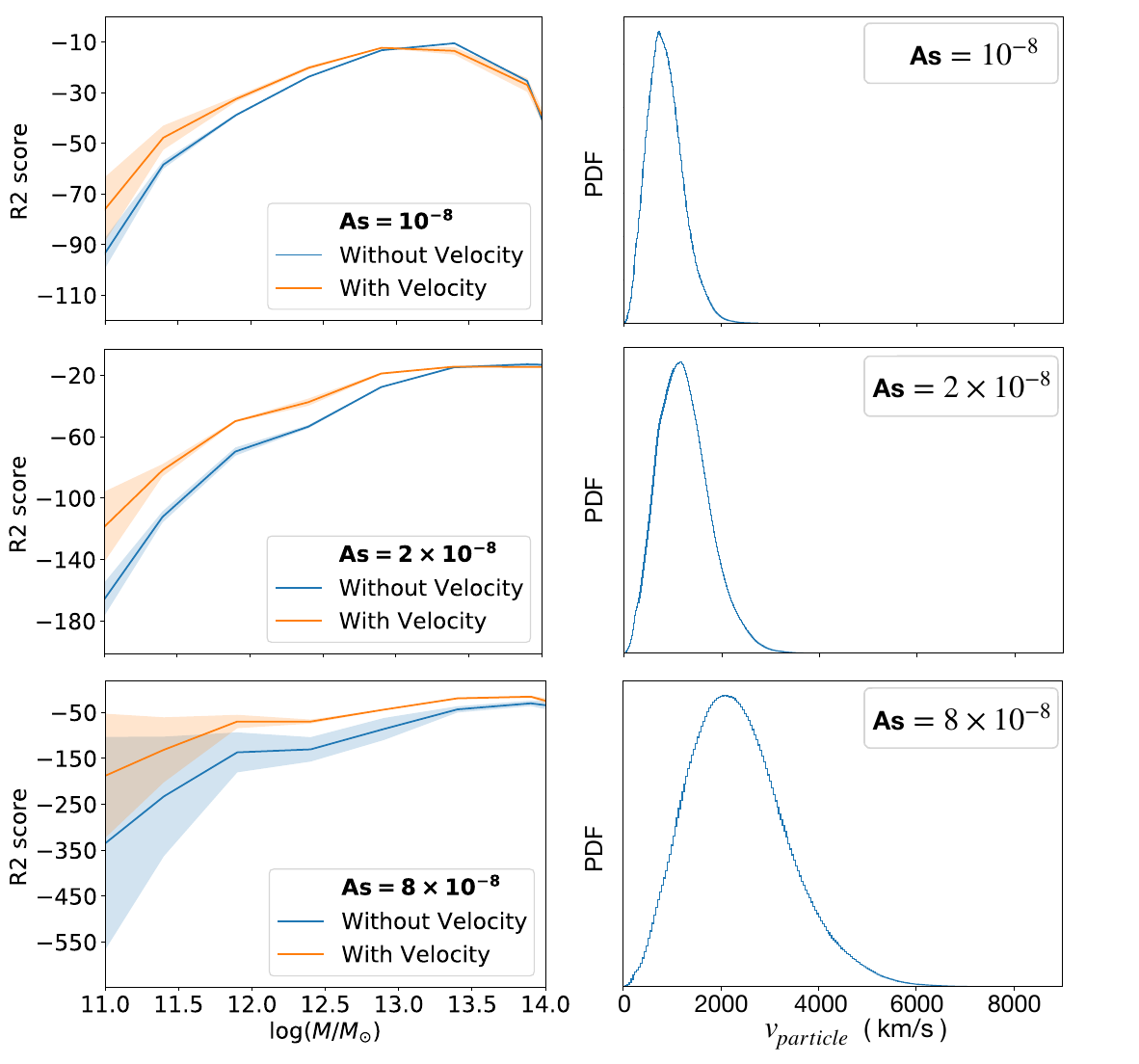}
\caption{{ The R2-score statistics is plotted versus mass for three different initial power spectrum in the left panel. In the right panel, we show the distribution of the dark matter particle's velocity. The right panel indicates that the peak of the velocity distribution of particles moves to larger values when we increase the initial curvature perturbations.}  }
\label{R2score}
\end{figure}

\begin{figure}
\centering
\includegraphics[scale=.5]{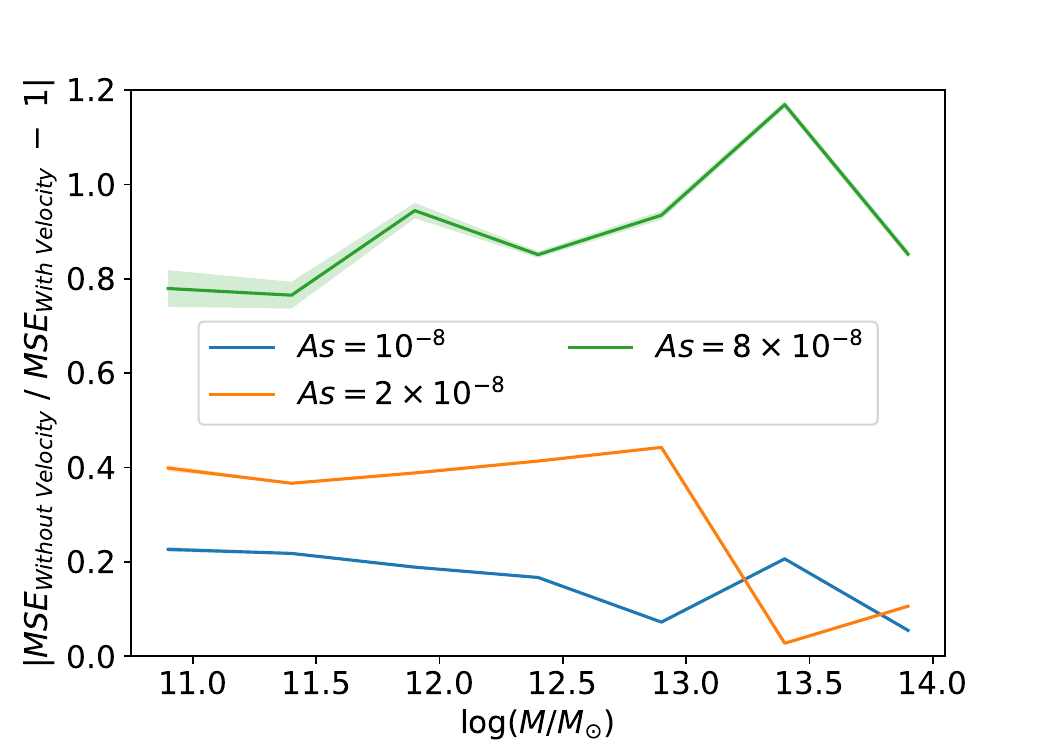}
\caption{The difference between the two (with and without velocity) models for different initial curvature perturbations are plotted versus mass. By increasing the power to $A_s = 8 \times 10^{-8}$, the difference between the two models increases to more than $80\%$. This implies that for higher $A_s$ values, the density field information is insufficient to predict the halo mass function accurately.}
\label{MSE}
\end{figure}

Again, we train our model five times for the new set of simulations with a higher initial power spectrum, and we compare the R2-score of bootstrapped cross-validated samples. The result is demonstrated in Fig.\ref{R2score}.
As before, the solid lines are the extrapolation of the mean of the R2-score in each mass bin, and the shaded regions are the 1$\sigma$ uncertainty obtained from bootstrap.
{{The right panel in Fig.\ref{R2score} indicates that by increasing the initial power spectrum, the distribution of the velocities shifts to higher values. This implies that by increasing the initial power spectrum, we can bring the effect of the velocity field to larger mass scales which effectively means that by looking at a fixed halo with a specific mass in this non-standard case, we are looking at smaller mass dark matter halos compared to the standard model case. On the other hand this interpretation is equivalent to the case, which we look at a halo in later times. The  corresponding initial redshift for $A_s= 10^{-8}, 2\times 10^{-8}, 8\times 10 ^{-8}$ are $z_{ini}\simeq 44,31,15$. }}The left panel of Fig.\ref{R2score} shows that the precision of the model without velocity field drops significantly when the initial power spectrum increases. Moreover, the difference between the model trained with velocity field information and the one without it increases noticeably when increasing the initial power. To show this more explicitly, we measure the difference between the two models for each case. As both models (with and without velocity channels) are trained on the same data set, the standard deviation in each mass bin is the same. Thus using the mean squared error (MSE) of each model should be sufficient to compare their performances. We measure the relative difference of the two models in each mass bin, $d(M_\text{bin})$, using:
\\
\begin{align*}
d(M_\text{bin}) = 
\frac{| \text{MSE}_\text{without}(M_\text{bin}) -  \text{MSE}_\text{with}(M_\text{bin}) |  }
{ \text{MSE}_\text{with}(M_\text{bin}) },
\end{align*}
\\
where,
\\
\begin{align*}
\text{MSE}(M_\text{bin}) = \frac{1}{N} \sum_{i=1}^{N} (
M_\text{pred}^{(i)} - M_\text{real}^{(i)} 
 )^2 ,
\end{align*}
\\
and $N$ is the total number of particles in that specific mass bin. Fig.\ref{MSE} shows the result of this distance for the different amplitude of initial curvature perturbations, where the errors are obtained from cross-validation.
{{The figure implies that the prediction of the model with velocity information is much better than the one without velocity for the large amplitude of initial curvature perturbations. This is consistent with our previous discussion that higher initial amplitudes bring the effect of velocities into larger mass halos compared to the $\Lambda\text{CDM}$ case. This is done by increasing the ratio  $\delta_{\text{ell}}/\delta_{\text{sc}}$ at larger mass halos to mock the effect of velocity shear field, which is visible in the small mass halos in the $\Lambda \text{CDM}$ paradigm.
Or, equivalently the large amplitude bring the initial box of simulation to lower redshifts.  This is done by looking at the Universe, which its growth function $D(z)$ is larger than the its value in $z_{ini}=100$.\\ 
As a result, we assert that the velocity information is vital on non-linear scales. Our result obtained from a collapse model independent deep learning framework illustrates that the ellipsoidal collapse model will diverge from the spherical collapse model on a small scale in the standard model. Using a top-down instead of a bottom-up view, we define the non-linear regime as a regime where the velocity field information will provide additional information to the one already existing in the density field to construct the halo mass function. In this sense, the Fig.\ref{MSE} asserts on our critical point providing how much non-linearity we can expect while increasing the $A_s$, and with the discussion in Sec.\ref{sec2}, how much non-linearity we should expect while moving to lower redshifts or smaller mass scales.}}

\section{Conclusions and Future Remarks}\label{sec6}
We developed an interpretable deep-learning framework to shed light on the process of structure formation. We train a CNN model on an initial simulation box at $z_{ini}=100$ to find the final destination halo mass for each DM particle. Our method, in comparison with its previous studies, has the advantage of being simple in terms of the initial features. We use our prediction for the halo mass function as the indicator of the performance of our machinery. Our results in terms of the halo mass function are in agreement with the one from the simulation box up to $1 \sigma$ in the halo mass range of $11 <\log(M/M_{\odot}) < 13$. 
We use our framework to find the importance of the velocity field in the structure formation process. We train our CNN on two different sets of samples: one on the images with velocity channels and the other without velocity channels. Our results imply that the effect of the velocity field in the standard model of cosmology given the initial curvature perturbations $A_s = 2\times10^{-9}$ and initial box $z_{ini}=100$ is negligible in the mass range of $10.5<\log(M/M_{\odot})< 14$. {{To study the effect of the velocity field on non-linear scales, we use a novel idea. We increase the initial curvature perturbations to increase the variance of perturbations in the scales that are in the scope of our simulation. This causes massive halos to mimic the effect of velocity fields of a small mass range.  Accordingly, the effect of velocity becomes visible in larger mass halos of our simulations with the given resolution. Equivalently, this means that we choose the initial box in lower redshifts (e.g. $A_s=8\times 10^{-8}$ corresponds to $z_{ini}\simeq 15$. We observe that increasing the initial curvature perturbations increases the non-linearities and velocity field in the structure formation process becomes important, correspondingly. \\
We discuss why in smaller scales or lower redshifts, the combination of both velocity and density fields will have more relevant information for our CNN map compared to the case with only the density field.
In this direction, we show that the normalized cross power-spectrum deviates from unity on non-linear scales. 

{{It is worth noting that we use the Zel’dovich approximation, based on linear perturbation theory, for the initial conditions in our simulations. Although this approximation assumes that velocity and density fields are fully correlated, as shown in Fig. \ref{fig:normalized-cross}, this correlation does not hold for unknown reasons when $A_s$ is large. Consequently, for large $A_s$, the velocity and density fields exhibit independent behavior. This decoupling explains why, in the $\Lambda$CDM case, adding velocity information does not improve the machine's accuracy, as high correlation is already present between density and velocity. However, when $A_s$ is larger, the inclusion of velocity information enhances accuracy. While this seems inconsistent with the assumptions of the Zel’dovich approximation, the observed decoupling of velocity and density fields suffices for the purpose of our paper. Nevertheless, for developing machine learning methods applicable to cosmological studies, a more consistent approach, such as using higher-order initial conditions, is essential.}}

By increasing the amplitude of the initial conditions the deviation from one occurs.
This result implies that the effect of the velocity field in structure formation is both scale-dependent and redshift dependent.
The variance of perturbation increased in smaller mass scale in a specific redshift. Equivalently, it gets larger in lower redshifts for a specific mass.\\
Our work shows the vast potential of Convolutional Neural Networks to extract higher-order information from basic phase space information and their capability to help us investigate fundamental physical questions due to their simplicity and interpretability. In another word, our results are more interesting when one notes that all these reasonable physical expectations re-obtained in our work are just obtained by analyzing the crude result of a CNN map trained on two simulation snapshots. This is another plus in favor of those who believe that ML black boxes can, in fact, have correct physical information rather than just being predictivity tools without physical insight.}}
It is noteworthy that the halo mass, in our analysis, is defined based on the number of bound particles and we do not have any pre-assumptions on the shape of halos. So, there is no constraint on our learning process to select any specific collapse model. {Our results show that incorporating velocity field information alongside the density field is essential for accurately studying the halo mass function in the non-linear regime, which occurs for large $A_s$ considerations. This finding underscores the shortcomings of the simple spherical collapse model, which proves inadequate under such conditions. Moreover, these results align with expectations driven by non-linear effects and were obtained without relying on any prior physical assumptions.}  Accordingly, it will be interesting to study the effect of different smoothing window functions on the halo mass prediction in a separate work.\\
Future work might study the correlation between the accuracy of halo mass prediction and the particle's distance from its host halo center. This is to integrate the idea of peak theory and excursion set theory of peaks in our ML methods. We can use our deep learning method to probe other cosmologies; for example, the models that velocity may play an important role in large mass halo formation. Also, this method can be used to investigate the effect of the velocity field on the structure formation in extensions of the standard model (e.g., massive neutrinos) and alternative models (e.g., $k$-essence models). \\\\


\section*{ACKNOWLEDGMENTS} 
We would like to thank the anonymous referee for their insightful comment  that elevated the manuscript to a whole new level.
We thank Francisco Villaescusa-Navarro for his helpful comment and discussion on the normalized cross-power spectrum. We also thank Arya Farahi for many useful discussions. The authors would like to especially thank Thomas Montandon and Julian Adamek for their valuable discussions on higher-order initial condition generators, as well as for their assistance in installing the MonofonIC code.\\
This work is supported by Oslo University computational facilities and a grant from
the Swiss National Supercomputing Centre (CSCS) under project ID s1051. \\
During part of this work, SE was supported by Perimeter Institute for Theoretical Physics. Research at Perimeter Institute is supported by the Government of Canada through Industry Canada and by the Province of Ontario through the Ministry of Research and Innovation. \\
SB is partially supported by the Abdus Salam International Center for Theoretical Physics (ICTP) under the regular associateship scheme. \\This research is supported by Sharif University of Technology Office of Vice President for Research under Grant No. G960202 and G4010204.\\ \\


\appendix
\section{Initial condition construction}

{{To generate the initial conditions for our simulations, we utilize the Zeldovich approximation, which is based on the first-order perturbation theory. Following the framework outlined in \cite{Adamek:2016zes}, we begin our simulations at a sufficiently high redshift $z = 100$, ensuring the validity of this approximation.
		
		The Zeldovich approximation relates the displacement of particles from their initial Lagrangian positions $\mathbf{q}$ to their final Eulerian positions $\mathbf{x}$ through the equation:
		
		\be
		\mathbf{x}(\mathbf{q}, \tau) = \mathbf{q} + \mathbf{\Psi}(\mathbf{q}, \tau),
		\ee
		
		where $\mathbf{\Psi}(\mathbf{q}, \tau)$ is the displacement field. The density perturbation $\delta$ and the velocity field $\mathbf{v}$ are derived from a single initial matter power spectrum $P(k)$ using the linear theory.  Consequently, the approximation suggests that the velocity and density fields contain the same underlying information, as they are both sourced from the same initial power spectrum.
		
		The fact that the initial velocity and density fields contain the same information suggests that incorporating velocity into machine learning methods may be redundant. However, as illustrated in Figure \ref{fig:normalized-cross}, we observe a decoupling between the density and velocity fields at large scales, when as we increase the amplitude of scalar fluctuations, $A_s$. 
		
		This behavior likely arises from numerical effects that remain poorly understood, particularly since we employed the Zeldovich approximation for generating initial conditions. Under this approximation, we expect a strong correlation between the initial density and velocity fields across all scales. However, our results show that at larger values of $A_s$, the velocity and density fields exhibit independent information. While it may seem counter-intuitive for velocity to be independent, this separation—due to either physical or numerical reasons—is particularly intriguing for our study. Our primary interest lies in determining whether incorporating independent velocity information can enhance the accuracy of our machine learning scheme.
		
		To enhance the performance of our machine learning approach for predicting halo mass functions, we propose two future directions. First, we could initialize our simulations at higher redshifts and provide snapshots at lower redshifts for training. This would allow the density and velocity fields to evolve, leading to distinct information content as they diverge from their initial conditions.
		
		Alternatively, we could utilize higher-order perturbation theory, such as the MonofonIC approach \cite{Montandon:2022ulz}, to generate the density, velocity, and potential fields. This method offers the potential for physically independent information across these fields, enriching the training dataset for our machine learning models and improving their predictive capabilities.}}

\end{document}